# Possibly novel demonstration of nonlocality for two maximally entangled particles

Demetrios Kalamidas

*Institute for Ultrafast Spectroscopy and Lasers, City College of the City University of New York, 138$^{th}$ Street & Convent Avenue, New York, NY 10031, USA*

Abstact

We present, at the gedanken level, a possibly novel non-statistical demonstration of nonlocality for two maximally entangled particles. The argument requires only two alternative experimental contexts, only one and the same single-particle observable, and leads to a contradiction with a particular form local realism in 50% of experimental runs. However, the argument only goes through for a plausible but not definitive notion of locality.

Inspired by Hardy's gedanken experiment [1], we present a non-statistical demonstration of nonlocality, again at the gedanken level, for two maximally entangled particles (Bell states). The argument requires only two alternative experimental contexts (instead of four) and only one and the same single-particle observable (instead of two different ones) for each of the two particles comprising the Bell states. However, as will be explained, the argument can only go through for a plausible but not definitive notion of locality.

Consider the two schemes, (a) and (b), of Fig.1. An electron **e**$^-$ and a positron **e**$^+$ are created by independent sources at distant locations and they arrive simultaneously at their respective interferometers, I$^-$ for **e**$^-$ and I$^+$ for **e**$^+$. In (a) there are two points of intersection, at P and Q, so that if the (**e**$^-$,**e**$^+$) pair occupies path combination {a$^-$, a$^+$} or {b$^-$, b$^+$} it annihilates into radiation, whose state is denoted by $|\gamma\rangle^x$ (the superscript x=P,Q indicating where the radiation originates). In (b) the two points of intersection are at R and S and annihilation occurs if the (**e⁻**,**e⁺**) pair occupies path combination {a⁻, b⁺} or {b⁻, a⁺}, the state of the radiation being $|\gamma\rangle^y$ (y=R,S). All of the beam splitters shown in the construction of the interferometers are 50/50.

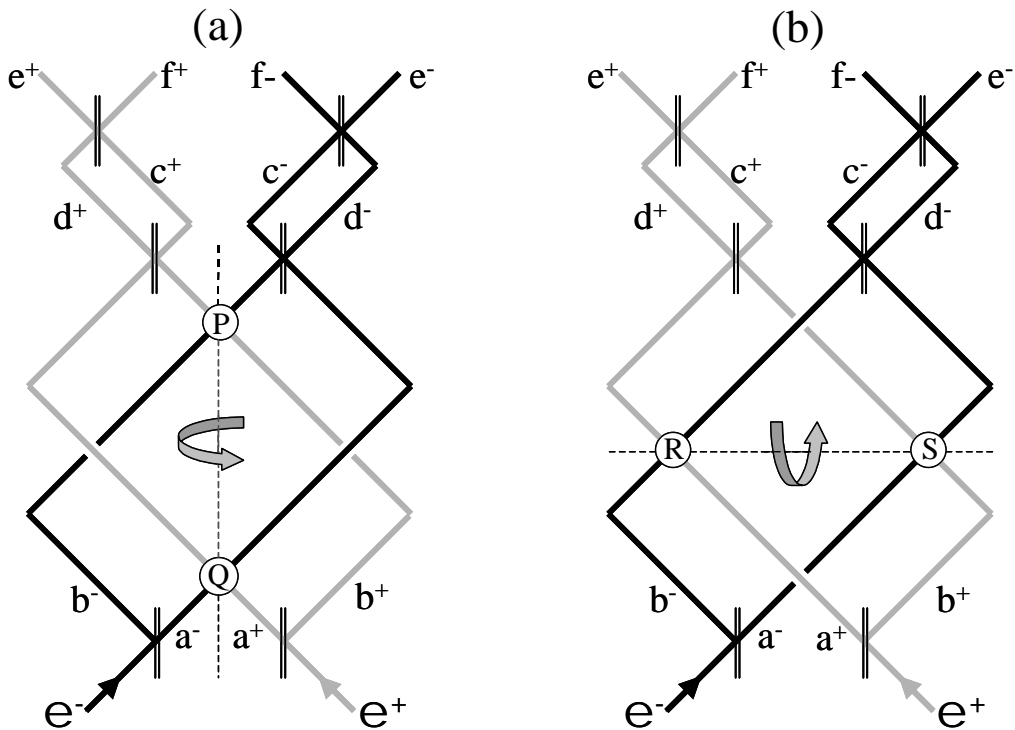

Fig.1. Intuitive depiction of the two interferometer arrangements, (a) and (b), used in the argument for nonlocality presented in the text. In (a), intersection only at points P and Q is obtained by having the plane of, say, I⁺ rotated slightly about axis PQ. In (b), intersection only at points R and S is obtained by having the plane of I⁺ rotated slightly about axis RS. Going from (a) to (b) involves only a slight change in the relative position of the two interferometers.

The action of the first pair of beam splitters on the input modes is given by

$$|IN^{\mp}\rangle \to \frac{1}{\sqrt{2}}(|a^{\mp}\rangle + i|b^{\mp}\rangle),$$

where the kets denote the presence of a particle in a particular path (i.e., $|a^-\rangle$ denotes the presence of an electron in path $a^-$). In analogous fashion, the subsequent two pairs of beam splitters transform their corresponding inputs as follows:

$$|a^{\mp}\rangle \to \frac{1}{\sqrt{2}}(|c^{\mp}\rangle + i|d^{\mp}\rangle), \quad |b^{\mp}\rangle \to \frac{1}{\sqrt{2}}(|d^{\mp}\rangle + i|c^{\mp}\rangle);$$

$$|c^{\mp}\rangle \to \frac{1}{\sqrt{2}}(|e^{\mp}\rangle + i|f^{\mp}\rangle), \quad |d^{\mp}\rangle \to \frac{1}{\sqrt{2}}(|f^{\mp}\rangle + i|e^{\mp}\rangle).$$

Taking into account the transformations stated thus far, the final state emerging in scheme (a) is given by

$$\frac{1}{2}|\gamma\rangle^Q - \frac{1}{2}|\gamma\rangle^P - \frac{i}{2}|e^-\rangle|f^+\rangle - \frac{i}{2}|f^-\rangle|e^+\rangle \tag{1}$$

while for scheme (b) the final state is given by

$$\frac{i}{2}|\gamma\rangle^R + \frac{i}{2}|\gamma\rangle^S + \frac{1}{2}|e^-\rangle|e^+\rangle - \frac{1}{2}|f^-\rangle|f^+\rangle. \tag{2}$$

We will now demonstrate how the interferometer arrangements described above lead to a contradiction between a particular form of local realism and certain predictions of quantum mechanics. First we prescribe what is meant by 'local realism' in the context of this gedanken experiment.

Adopting realism, we assume that the (**e⁻**,**e⁺**) pair is completely described by two sets of hidden variables, $\{\lambda\}$ for **e⁻** and $\{\mu\}$ for **e⁺**, since the two particles are created by independent sources. The elements of $\{\lambda\}$ and $\{\mu\}$ can take on different values for each run of the experiment (in other words, for each (**e⁻**,**e⁺**) pair taken from an ensemble of what quantum mechanics regards as identically prepared states). Lets further assert that the outcomes of any single-particle measurements on the (**e⁻**,**e⁺**) pair have been predetermined by the values within $\{\lambda\}$ and $\{\mu\}$ in the following manner: For measurements on **e⁻**, the outcomes have been predetermined by the values within $\{\lambda\}$ subject only to the *intrinsic properties* of interferometer $I^-$; for measurements on **e⁺**, the outcomes have been predetermined by the values within $\{\mu\}$ subject only to the intrinsic properties of interferometer $I^+$. The 'intrinsic properties' of each interferometer are the relative phases between paths and the reflectance/transmittance of each of the beam splitters it contains, nothing more. We denote the phase relations in $I^-,I^+$ by $\varphi(I^\mp) \equiv [\varphi(a^\mp,b^\mp);\varphi(c^\mp,d^\mp)]$, where $\varphi(x^\mp, y^\mp)$ means "the relative phase between paths x and y". For the beam splitters we write $\{BS^\mp\}$, which denotes the set of beam splitters, each with its particular reflectance/transmittance, contained in $I^-,I^+$. In doing this we have introduced a *plausible* notion of locality because each particle's measurement outcome is determined only by the hidden variables that were defined at its source, subject only to the 'local settings' (intrinsic properties) it encountered during its propagation through its interferometer, and by the fact that a measurement outcome in itself implies that the particle survived annihilation and therefore never interacted with its partner at any of the spatially localized regions (points) P,Q or R,S. This notion of locality is, at best, 'plausible' and not definitive because the interferometer arrangements considered here do not provide space-like separation between **e⁻** and **e⁺** at times before measurements occur.

Now suppose there is a detector in each of the four output paths e-, f-, e⁺, f⁺. If, for instance, we consider scheme (a) and an electron is detected in path e-, we denote the event by $E^-_{\{\lambda\}}[\varphi(I^-),\{BS^-\}]=1$; non-detection in that path is denoted by $E^-_{\{\lambda\}}[\varphi(I^-),\{BS^-\}]=0$. With this notation the dependence of an outcome on the hidden variables, subject to the intrinsic properties of the associated interferometer, is made explicit. Analogous notation is used for the rest of the detectors.

Applying the form of local realism described above leads to the following statements involving detection outcomes:

For the interferometer arrangement of scheme (a) we have

$$E^-_{\{\lambda\}}[\varphi(I^-),\{BS^-\}]E^+_{\{\mu\}}[\varphi(I^+),\{BS^+\}]=0 \qquad (3a)$$

and

$$F^-_{\{\lambda\}}[\varphi(I^-),\{BS^-\}]F^+_{\{\mu\}}[\varphi(I^+),\{BS^+\}]=0 \qquad (3b)$$

for *all* runs of the experiment because there are no $|e^-\rangle|e^+\rangle$ or $|f^-\rangle|f^+\rangle$ terms present in (1).

As shown in Fig.1, the interferometer arrangement of scheme (b) involves a different relative positioning of I⁻ and I⁺ with respect to that of scheme (a) and may be obtained by, say, a *slight* tilting of the plane of I⁺.

Yet, in going from (a) to (b), the intrinsic properties of the interferometers have not changed and neither have the particle sources undergone any modification. Therefore, had we performed the experiment of scheme (b) we would have

$$E^{-}_{\{\lambda\}}[\varphi(I^{-}),\{BS^{-}\}]E^{+}_{\{\mu\}}[\varphi(I^{+}),\{BS^{+}\}]=1 \qquad (4a)$$

in 25% of runs because of the term $\frac{1}{2}|e^{-}\rangle|e^{+}\rangle$ present in (2) and

$$F^{-}_{\{\lambda\}}[\varphi(I^{-}),\{BS^{-}\}]F^{+}_{\{\mu\}}[\varphi(I^{+}),\{BS^{+}\}]=1 \qquad (4b)$$

in 25% of runs because of the term $\frac{1}{2}|f^{-}\rangle|f^{+}\rangle$ also present in (2). However, *both* of the joint outcomes expressed by (4a) and (4b) directly contradict (3a) and (3b), respectively, and therefore a contradiction with the given local realistic description of the two interferometer arrangements occurs for 50% of runs in the experiment of scheme (b). Of course, as already mentioned, this conclusion is based on the restriction that no causal influences are exchanged between **e⁻** and **e⁺** with the ability to deterministically change the outcome of a single-particle measurement as a result of the slight difference in the spatial positioning of $I^{-},I^{+}$ in going from (a) to (b).

Upon inspection of (1) we can observe that the (**e⁻**,**e⁺**) pair is in a *maximally path-entangled state* since $-\frac{i}{2}|e^{-}\rangle|f^{+}\rangle-\frac{i}{2}|f^{-}\rangle|e^{+}\rangle=-\frac{i}{\sqrt{2}}|\Psi^{(+)}\rangle$, where $|\Psi^{(+)}\rangle=\frac{1}{\sqrt{2}}(|e^{-}\rangle|f^{+}\rangle+|f^{-}\rangle|e^{+}\rangle)$ is a Bell state. Similarly, the state in (2) is $\frac{1}{\sqrt{2}}|\Phi^{(-)}\rangle$, where $|\Phi^{(-)}\rangle=\frac{1}{\sqrt{2}}(|e^{-}\rangle|e^{+}\rangle-|f^{-}\rangle|f^{+}\rangle)$ is another Bell state. So, the contradiction between {(3a),(3b)} and {(4a),(4b)} demonstrates that a non-

statistical proof of nonlocality *may* be possible for Bell states, contrary to the view expressed in [2] and [3].

The novel features of this gedanken experiment are its reliance on only two alternative contexts and the fact that an analysis of measurement outcomes of two different single-particle observables is not necessary: the intrinsic properties of each interferometer do *not* change when going from (a) to (b), the same observable is examined for both particles in both experimental contexts. Our argument differs with that developed by Hardy [1,2] because he considers alternative choices of space-like separated single-particle observables for one and the same prepared state whereas we propose a *possibly* nonlocal preparation of two different Bell states. The purported 'nonlocal preparation' of the two different Bell states used in our argument is plausible, at best, since it is based on the exclusion of the aforementioned exchange of causal influences between the two particles. However, if they exist, such causal influences would have to possess extraordinary attributes: able to remain undetected to this date, not diminishing with distance, unaffected by any line-of-sight material obstruction, and able to cause dramatic deterministic changes from one set-up to another even though the spatial configurations differ by a slight amount.

Finally, in addition to being constrained by the particular notion of locality we have adopted, the non-statistical demonstration of nonlocality for Bell states given here is valid only for the ideal conditions (perfect sources, perfect alignments, perfect detectors, etc.) presumed to hold in the gedanken realm. If a feasible realization of this gedanken experiment is possible, it would most likely involve entangled photons and would require the formulation of an appropriate inequality to account for the non-ideal conditions.

An attempt was made by Wu et al. [4] to demonstrate a Hardy-type violation of local realism with maximally entangled two-photon states. However, their claims have been criticized by Cereceda [5,6] and Cabello [7] who note that their proof was not based on dichotomic single-particle observables as were Hardy's arguments [1,2]. If it were possible to rigorously exclude the existence of the 'extraordinary' causal influences we have described then the demonstration of nonlocality presented in this Letter would, to our knowledge, exhibit the strongest contradiction (50%) between local realism and quantum mechanics for Bell states.

Partial financial support for this research was provided by DoD.

[1] L. Hardy, Phys. Rev. Lett. 68 (1992) 2981.
[2] L. Hardy, Phys. Rev. Lett. 71 (1993) 1665.
[3] S. Goldstein, Phys. Rev. Lett. 72 (1994) 1951.
[4] X. Wu et al., Phys. Rev. A 53 (1996) R1927.
[5] J.L. Cereceda, Phys. Rev. A 55 (1997) 3968.
[6] J. L. Cereceda, Phys. Lett. A 263 (1999) 232.
[7] A. Cabello, Phys. Rev. A 61 (2000) 022119.